\documentclass[12pt, a4paper]{article}
\pdfoutput=1
\usepackage[dvipdfmx]{graphicx}
\usepackage{amssymb}
\usepackage{amsmath}
\usepackage{bm}
\usepackage{color}
\usepackage{cite}
\usepackage{slashed}
\usepackage{subfigure}
\usepackage{epstopdf}            
\usepackage{epsfig}
            
\newcommand{\beq}{\begin{equation}}
\newcommand{\eeq}{\end{equation}}

\newcommand{\lv}{{L^\prime}}

\usepackage[colorlinks=true, linkcolor=black, citecolor=black,
urlcolor=black]{hyperref}

\begin{document}

\begin{titlepage}

\begin{flushright}
 \begin{minipage}{0.2\linewidth}
  \normalsize
  UT-17-22 \\*[50pt]
 \end{minipage}
\end{flushright}

\begin{center}

{\Large
{\bf 
{Interplay between the $b \to s ll$ anomalies \\ and dark matter physics}
}
}

\vskip 2cm

Junichiro Kawamura$^{1}$, 
Shohei Okawa$^2$
and
Yuji Omura$^{3}$

\vskip 0.5cm

{\it $^1$Department of Physics, University of Tokyo, 
Tokyo 113-0033, Japan
}\\[3pt]
{\it $^2$
Department of Physics, Nagoya University, Nagoya 464-8602, Japan}\\[3pt]
{\it $^3$
Kobayashi-Maskawa Institute for the Origin of Particles and the
Universe, \\ Nagoya University, Nagoya 464-8602, Japan}\\[3pt]

\vskip 1.5cm

\begin{abstract}
Recently, the LHCb collaboration has reported the excesses in the $b \to s ll$ processes.
One of the promising candidates for new physics to explain the anomalies is 
the extended Standard Model (SM) with vector-like quarks and leptons.
In that model, Yukawa couplings between the extra fermions and SM fermions
are introduced, adding extra scalars. Then, the box diagrams involving 
the extra fields achieve the $b \to s ll$ anomalies.
It has been known that the excesses require the large Yukawa couplings of leptons, 
so that this kind of model can be tested by studying correlations with other observables. 
In this paper, we consider the extra scalar to be a dark matter (DM) candidate,
and investigate DM physics as well as the flavor physics and the LHC physics. 
The DM relic density and the direct-detection cross section are also dominantly given by the Yukawa couplings,
so that we find some explicit correlations between DM physics and the flavor physics.
In particular, we find the predictions of the $b \to s ll$ anomalies against the direct detection of DM.  
\end{abstract}

\end{center}
\end{titlepage}

\section{Introduction}
Recently, the LHCb collaboration has reported that there are deviations from the
Standard Model (SM) predictions in the $b \to s ll$ processes.
In the experiment, 
the branching fractions of $B \to K^{(*)} \, ll$ ($l=e, \, \mu$) are measured, 
and lepton universalities and angular distributions are studied in each process. 
One excess is reported in the ratio 
between BR($B^+ \to K^+ \, \mu \mu$) and BR($B^+ \to K^+ \,ee$) 
in the region with $1$ GeV$^2$ $\leq q^2 \leq 6$ GeV$^2$, 
where $q^2$ is the invariant mass of two leptons in the final state \cite{Aaij:2014ora}. 
The experimental result suggests the smaller value of  BR($B^+ \to K^+ \, \mu \mu$)
than the SM prediction, and the deviation is about $2.6 \sigma$~\cite{Aaij:2014ora}.  
Recently, a similar deviation is discovered in $B \to K^* \, \mu \mu$ \cite{LHCbnew}.  
The $B$ decay to $\mu$ pair in this process is again smaller than the SM prediction.
Similar indications are also reported in $B \to \phi \, \mu \mu$ \cite{Aaij:2015esa} and $\Lambda_b \to \Lambda \, \mu \mu$ \cite{Aaij:2015xza} in the same $q^2$ region. 
Moreover, the disagreement between the experimental results
and the SM prediction of the angular distribution in $B \to K^* \, \mu \mu$
is also one of the longstanding issues \cite{Aaij:2013qta,Aaij:2015oid}. The CMS collaboration has shown
the result that may be consistent with the SM prediction, but the deviation is still 
large in the LHCb experiment and the others.
Thus, there might be some issues in the $b \to s$ transition associated with $\mu$. 

The SM predicts that namely $C_7$, $C_9$ and $C_{10}$ operators contribute to the $b \to s ll$ processes. 
$C_7$ can not give sizable contributions to the processes, 
because it corresponds to the electric dipole operator that
is strictly constrained by $B \to X_s \gamma$. 
In the region with $1$ GeV$^2$ $\leq q^2 \leq 6$ GeV$^2$, the $C_9$ and $C_{10}$ operators  dominantly 
contribute to the branching ratios. The Wilson coefficients, $C^l_9$ and $C^l_{10}$, are defined as follows:
\begin{equation}
{\cal H}_\textrm{eff}= -\frac{4 G_F}{\sqrt{2}} V_{tb} V^*_{ts} \frac{e^2}{16 \pi^2}
\left  \{ C_9^{l}(\overline{s_L} \gamma_\mu b_L)
(\overline{l}\gamma^\mu l )
+C_{10}^{l}(\overline{s_L} \gamma_\mu b_L)
(\overline{l}\gamma^\mu \gamma_5 l) +h.c. \right \}.
\end{equation}
In the SM, $C^l_9$ and $C^l_{10}$ are almost flavor universal and the sizes are estimated as
$(C_9^{l})_{{\rm SM}} \simeq -(C_{10}^{l})_{{\rm SM}} \simeq 4$ at the bottom quark mass scale.
There is an ambiguity in the long-distance contribution \cite{Khodjamirian:2010vf}, which is expected to be ${\cal O}(\Lambda_{QCD}/m_b)$,
but the excesses seem to require much larger contributions to $C^l_9$ and $C^l_{10}$:
$(\Delta C_9^{l})/(C_9^{l})_{{\rm SM}} \simeq -0.2$ and $(\Delta C_{10}^{l})/(C_{10}^{l})_{{\rm SM}} \simeq 0.2$,
according to the global fitting \cite{Descotes-Genon:2013wba,Hiller:2014yaa,Altmannshofer:2015sma,Descotes-Genon:2015uva,Hurth:2016fbr,Altmannshofer:2017yso,DAmico:2017mtc,Geng:2017svp,Ciuchini:2017mik,Alok:2017jaf}. 
Here, $\Delta C_{9,10}^{l}$ denote the new physics contributions 
for $C^l_{9,10}$, respectively.

Many new physics scenarios have been proposed 
in order to explain these $b \to s ll$ anomalies. One simple way is to
introduce an extra gauged flavor symmetry \cite{Zprime}. 
In such a model, the non-vanishing charges are assigned to both $\mu$ and quarks 
and the extra gauge boson contributes to $\Delta C_{9,10}^{l}$ at the tree-level. 
Although the $B_s$ meson mixing strongly constrains the contributions, 
the large $\Delta C_{9,10}^{l}$ can be achieved successfully 
if the charge of $\mu$ is not vanishing in the mass base. 
Another candidate for these excesses is namely $leptoquark$ 
that is a scalar or vector field charged under $SU(3)_c$ \cite{leptoquark}.
The new field couples both leptons and quarks, 
so that the tree-level exchanging can generate $\Delta C_{9,10}^{l}$.

We can discuss the other possibility that the large $\Delta C_{9,10}^{l}$ is realized
by the one-loop diagrams involving extra fields \cite{DAmico:2017mtc,Gripaios:2015gra,Arnan:2016cpy,Belanger:2015nma,Hu:2016gpe,Poh:2017tfo,Kamenik:2017tnu}.
For instance, extra vector-like quarks and leptons can be introduced 
without suffering from gauge anomaly.
Further, Yukawa couplings between the SM fermions and the extra fields 
can be written down by introducing extra scalar fields.

There are several variations of the extra fields as discussed in Refs. \cite{Gripaios:2015gra,Arnan:2016cpy}. 
In all cases, ${\cal O}(1)$ Yukawa couplings of $\mu$ and extra leptons are required, 
while moderately small Yukawa couplings of quarks and extra quarks are necessary 
to realize large enough $\Delta C_{9,10}^{l}$ 
but to evade from the stringent bound on the $B_s$ meson mixing.
The masses of the extra fields can not be so large 
to explain such large $\Delta C_{9,10}^{l}$: 
the masses of the extra fields 
should be in the range between ${\cal O}(100)$ GeV and ${\cal O}(1)$ TeV. 
Thus it is inevitable to 
investigate the consistency of the setup with the other observables, 
such as the direct signals at the LHC and
the other flavor violating processes.

It is very interesting that the neutral fields among the extra fermions and scalars become
good dark matter (DM) candidates \cite{Gripaios:2015gra,Arnan:2016cpy,Belanger:2015nma}. The mass region favored by the excesses corresponds to
the one discussed in the Weakly Interacting Massive Particle (WIMP) DM scenario.
The size of the interaction with the SM fermions via the Yukawa couplings
would not be too large to achieve the DM relic density 
via the well-known thermal process~\cite{Belanger:2015nma,Okawa,Bhattacharya:2015xha}.
As pointed out in Ref. \cite{Okawa}, however, this kind of model faces the
stringent constraint from the direct-detection experiments of DM. 
The cross sections are almost on the upper bound 
of the latest LUX and XENON1T experiments \cite{LUX2015,LUX2016,XENON1T} 
if the observed DM relic density is explained by the thermal process. 
The scattering processes are induced 
by the photon and Z-boson exchanging at the one-loop level,
and the contributions cancel each other in some parameter region.
In addition, the tree-level diagrams involving the extra fermions 
become sizable depending on the alignment of the Yukawa couplings.  
Moreover, 
the direct searches for extra quarks and leptons at the LHC are well developed recently.  
Therefore, the careful integrated study is required to discuss the excesses 
at the LHCb in this kind of model.

Accordingly, 
we especially study the DM physics and the LHC physics in this extended model with extra fermions and scalars which couple to both quarks and leptons. 
The rest of this paper is organized as follows. 
We introduce our setup in Sec. \ref{sec2}, motivated by the $b \to s ll$ anomalies. 
We discuss direct searches for extra fermions at the LHC in Sec. \ref{sec3} 
and flavor phenomenology in Sec. \ref{sec4}. 
Then we discuss impacts on DM physics 
from the other observations, especially the $b \to s ll$ anomalies in Sec. \ref{sec5}.  
Based on the study of the DM and the LHC physics, 
we could obtain the upper limits on $\Delta C_{9,10}^{l}$.
We conclude our discussion in Sec. \ref{sec7}.

\section{Setup}
\label{sec2}
First of all, we introduce our model motivated 
by the $b \to s ll$ anomalies at the LHCb experiment.
We introduce an extra SU(2)$_L$-doublet quark, denoted by $Q^\prime$, 
and an extra SU(2)$_L$-doublet lepton, denoted by $\lv$, 
to enhance $C^l_9$ and $C^l_{10}$.
Their SM charges are summarized in Table \ref{table1}. 
\begin{table}[thb]
\begin{center}
\begin{tabular}{cccccc}
\hline
\hline
Fields & ~~spin~~   & ~~$\text{SU}(3)_c$~~ & ~~$\text{SU}(2)_L$~~  & ~~$\text{U}(1)_Y$~~ &~~U(1)$_{X}$~~     \\ \hline  
   $Q^\prime$  & 1/2 & ${\bf 3}$        &${\bf2}$      &         $1/6$      & $1$          \\
   $\lv$  & 1/2 & ${\bf 1}$        &${\bf2}$      &         $-1/2$      & $1$          \\
   $ X$  & 0 & ${\bf1}$         &${\bf 1}$      &            $0$   & $-1$         \\ \hline \hline
\end{tabular}
\end{center}
\caption{Extra fields in our model with global U(1)$_{X}$. }
\label{table1}
\end{table} 
We assign global U(1)$_{X}$ charges to $Q^\prime$ and $L^\prime$ 
to distinguish them from the SM quarks and leptons.
Note that all of the SM particles are neutral under the U(1)$_X$. 
We also introduce a complex scalar $X$ as a candidate for DM. 
The DM $X$ is charged under the global U(1)$_{X}$ symmetry, 
and it becomes stable if $X$ is lighter than both $Q^\prime$ and $L^\prime$.
The charge assignment of $X$ is shown in Table \ref{table1}.
Note that $L^\prime$ consists of charged and neutral fermions, 
and the neutral component possibly becomes a good DM candidate. 
We concentrate on the case that $X$ is DM in this paper. 

We can write down the potential for the fermions and the scalar: 

\begin{eqnarray}
V&=&V_{\rm F}+V_{\rm S}, \label{eq;Vall} \\
V_{\rm F}&=& m_{Q^\prime} \overline{Q^\prime}_LQ^\prime_R +m_{\lv} \overline{\lv}_L\lv_R + \lambda^q_{ i} \overline{Q^\prime}_R  X^\dagger Q^i_{L} + \lambda^l_{ i} \overline{\lv_R}  X^\dagger l^i_{L} + h.c., \label{eq;VB}  \\
V_{\rm S}&=& m^2_X |X|^2+ \lambda_H |X|^2 |H|^2 + \lambda_X |X|^4 -m^2_H |H|^2+ \lambda |H|^4. \label{eq;VX} 
\end{eqnarray}
In our notation, $(Q^1, \, Q^2, \, Q^3)$ correspond to $((V^\dagger_{1j} u^j_L,d_L)^T, \, (V^\dagger_{2j} u^j_L,s_L)^T, \, (V^\dagger_{3j} u^j_L,b_L)^T)$. $V_{ij}$ denotes the CKM matrix.
The charged components of $(l_L^1, \, l_L^2, \, l_L^3)$ correspond to $(e_L, \, \mu_L, \, \tau_L)$, respectively.
Each SM fermion in $Q^i$ and $l_L^i$ is the mass eigenstate.
$ \lambda^q_{ i}$ and $\lambda^l_i$ is the Yukawa coupling 
between the DM and the extra fermions.
The SM down-type quarks in $Q^i$ and the charged leptons in $l_L^i$ 
are the mass eigenstates, then 
we simply denote the Yukawa couplings as $ (\lambda^q_{ 1},\,\lambda^q_{ 2},\,\lambda^q_{ 3}) \equiv (\lambda_{d},\,\lambda_{ s},\,\lambda_{ b})$ and $ (\lambda^l_{ 1},\,\lambda^l_{ 2},\,\lambda^l_{ 3}) \equiv (\lambda_{e},\,\lambda_{\mu},\,\lambda_{ \tau})$. 
Note that the scalar potential $V_{\rm S}$ includes the SM Higgs boson, 
denoted by $H$.
There are many possibilities of the SM charges and spins of the extra fields.
If $X$ is charged under the electroweak symmetry as discussed 
in Refs.~\cite{Gripaios:2015gra,Arnan:2016cpy},
$Z$-boson exchanging diagram would lead too large cross section for the direct detection of DM.
In the case that $X$ is fermion and $Q^\prime$, $\lv$ are scalars, 
the relic density is drastically reduced because of the $s$-wave contribution.
Then, large Yukawa couplings, which are favored by the LHCb excesses, would predict 
too small relic DM abundance thermally.
The fermionic $X$ is, moreover, strongly constrained by the indirect detection of DM.

Based on this setup summarized in Table \ref{table1} 
and Eqs.~(\ref{eq;Vall})-(\ref{eq;VX}),  
we study phenomenology in the LHC, flavor and DM experiments.
First, we discuss constraints on these extra fermions and DM 
from the direct searches at the the LHC in Sec. \ref{sec3}. 
Then, we investigate the flavor physics and 
the correlation between the $b \to s ll$ anomalies and the DM physics 
in Sec. \ref{sec4} and Sec. \ref{sec5}, respectively.

\section{Constraint from the direct search at the LHC }
\label{sec3}
\newcommand{\MET}{E_T^{\rm miss}} 
The extra particles summarized in Table \ref{table1} could be
directly discovered at the LHC.
The extra fermion decays to a SM-singlet $X$ and a SM fermion via the Yukawa couplings.
The branching ratio of the extra fermion depends on the alignment 
of the Yukawa couplings.
Motivated by the $b \to s ll$ anomalies, we assume $|\lambda_\mu| \gg |\lambda_{e,\tau}|$, 
and then the extra charged lepton dominantly decays to a singlet and a muon.
The expected signal events have two energetic muons 
and large missing energy, $\mu\mu + \MET $. 
Note that the neutral extra lepton decays to a singlet and a neutrino, 
and the decay is totally invisible.

The extra up-type (down-type) quark decays to a singlet 
and a SM up-type (down-type) quark depending on the Yukawa couplings. 
If $\lambda_d$ is relatively large, the contribution of the extra quark exchanging diagram via the Yukawa coupling
becomes too large to evade the strong bound from the DM direct detection experiments.
On the other hand, the anomalies reported by the LHCb collaboration require
sizable $|\lambda_s \lambda_b|$, as discussed in Sec. \ref{sec4}.
Therefore, we assume $|\lambda_s \lambda_b| = 0.15$, that corresponds to the upper bound from the
$B_s$-$\overline{B_s}$ mixing (see Sec. \ref{sec4}),
and consider two cases, 
\begin{eqnarray}
(\rm A) && \lambda_b = 1.0, \,  \lambda_s=0.15,   \nonumber \\
(\rm B) && \lambda_b = \lambda_s = \sqrt{0.15}.      \nonumber
\end{eqnarray}
In both cases, $\lambda_d$ is assumed to be negligibly small. 
In the case (A), we expect that the extra quark 
dominantly decays to a singlet and a third-generation quark,  
and the expected signals are $tt+\MET$ or $bb+\MET$.   
While the extra quark decays into  both a third-generation quark 
or a second-generation quark with a almost same probability in the case (B).

Since any extra fermions decay to a singlet 
and produce  large missing energies, 
the extra fermions give similar signals to supersymmetric particles. 
We refer to the LHC results obtained 
in the analysis searching for charginos~\cite{llMET} 
to constrain the extra leptons. 
For the extra quarks, 
we refer to the experimental limits from 
$bb+\MET$~\cite{bbMET} and $jj + \MET$~\cite{jjMET} channels. 
We do not consider limits from $tt + \MET$ searches 
because the limits will be weaker than the limit from $bb+\MET$. 
The mono-jet search is also not important one in our model 
if the singlet $X$ is the DM. 
This is because the direct detection easily 
excludes mass degenerate region  
where the mono-jet search becomes potentially important, as discussed in Sec. \ref{sec5}.

We calculate the expected number of signals in each signal region
using MadGraph5~\cite{Alwall:2014hca}. 
The UFO format~\cite{Degrande:2011ua} model files 
are created by FeynRules~\cite{Alloul:2013bka}. 
The signal processes are generated from matrix elements 
with up to an extra parton, 
then these are passed to PYTHIA 6~\cite{Sjostrand:2006za}
for decaying top quarks, parton showering and hadronization.  
The MLM scheme~\cite{Caravaglios:1998yr} is used 
to match the matrix element and parton showering. 
The generated events are interfaced to DELPHES3~\cite{deFavereau:2013fsa} 
for fast detector simulation. 
The generated hadrons are clustered 
using the anti-$k_T$ algorithm~\cite{Cacciari:2008gp} 
with the radius parameter $\Delta R = 0.4$. 
We basically use the ATLAS DELPHES card, 
but efficiency formulas of muon reconstruction and b-tagging are 
changed based on Ref.~\cite{muon} and Ref.~\cite{btag}.  
The muon reconstruction efficiency is set to 
$0.7\ (|\eta|<0.1),\ 0.99\ (0.1<|\eta|<2.5)$ 
and 
the b-tagging efficiency obeys 
$1.3 \tanh{(0.002 p_T)} 30/(1+0.086p_T)$, 
where $\eta,\ p_T$ are pseudorapidity and transverse momentum of a jet, 
respectively.  
The generated events are normalized 
according to their cross sections obtained by the event generation 
and the integrated luminosity of the data:   
36.1 $\text{fb}^{-1}$ for $bb+\MET$ and $jj+\MET$ searches, 
and 13.3 $\text{fb}^{-1}$ for $ll +\MET$ search.

\begin{figure}[!t]
\centering 
{\epsfig{figure=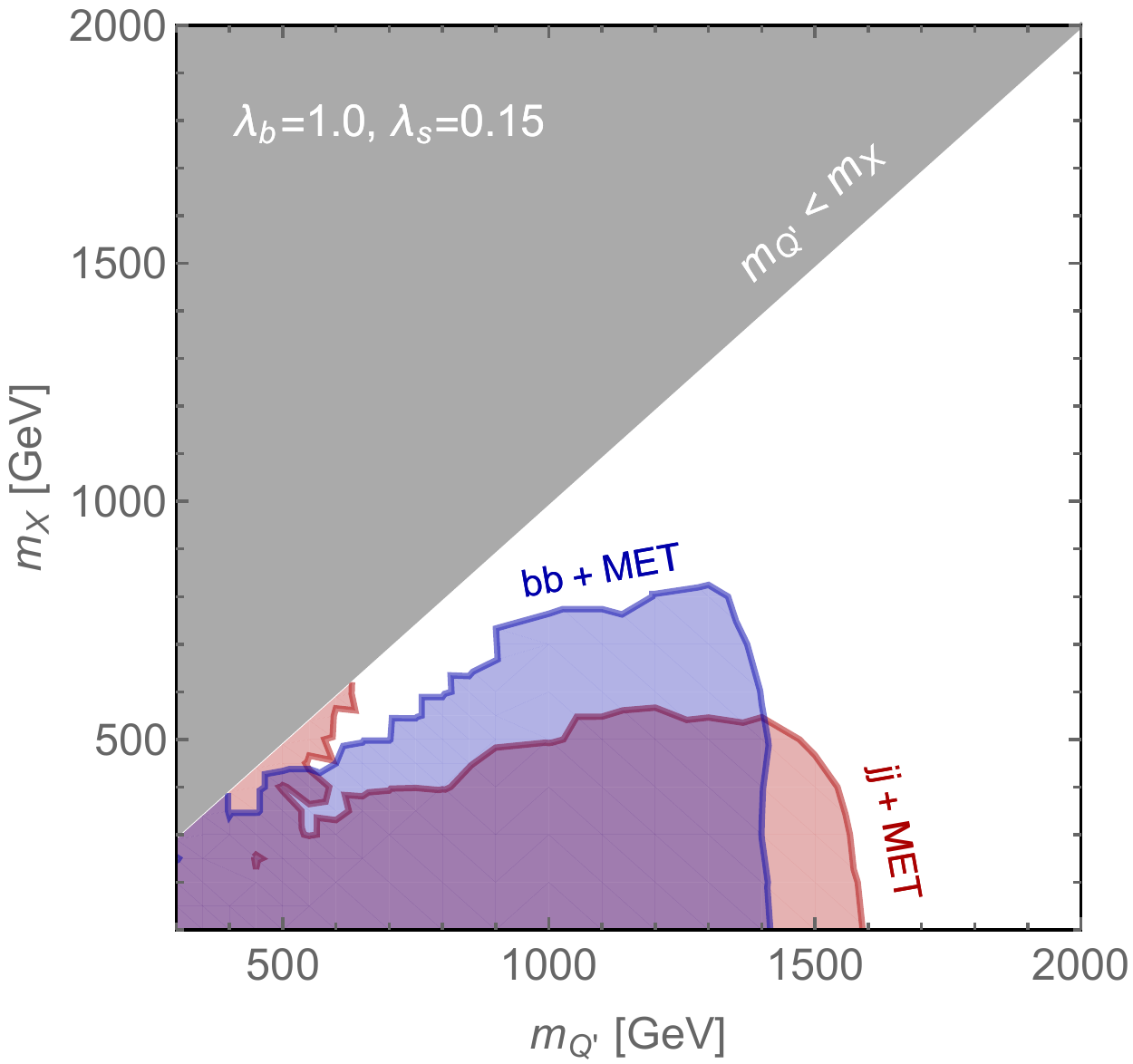,width=0.45\textwidth}} 
{\epsfig{figure=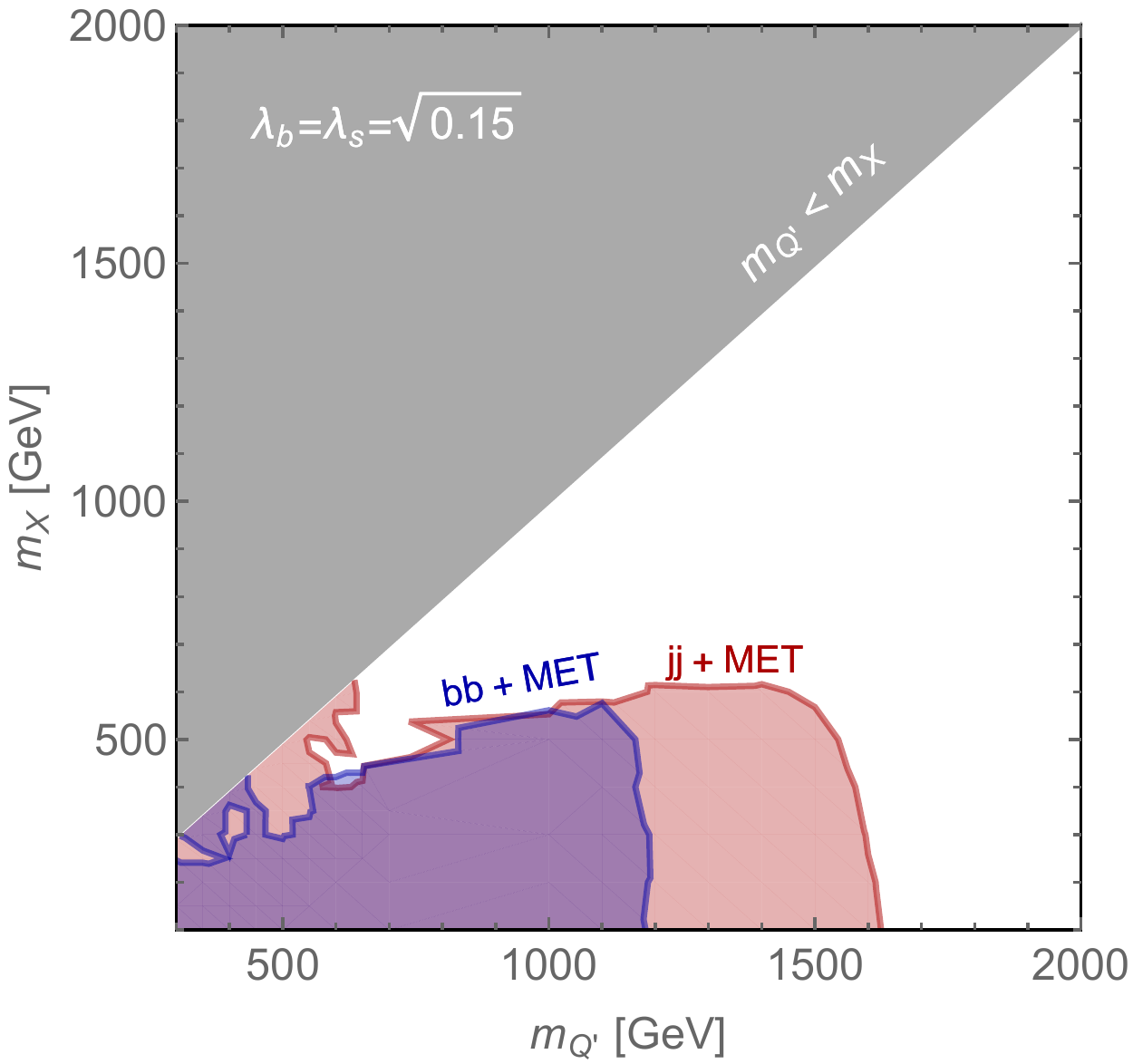,width=0.45\textwidth}} 
\caption{LHC bound for case(A) (left) and case (B) (right). 
The red regions are excluded by the $jj+\MET$ search and 
the blue regions are excluded by the $bb+\MET$ search. } 
\label{fig:LHCquark}
\end{figure}
Figure~\ref{fig:LHCquark} shows exclusion limits on the extra quarks
from the latest searches for $bb+\MET$ (blue) and $jj+\MET$ (red) events.  
We see in the case (A) 
that the $jj + \MET$ search is sensitive to 
heavy extra quark region with the light singlet 
while the $bb+\MET$ search is more sensitive 
to the heavier singlet region. 
In the case (B), 
the limits from $bb+\MET$ is significantly weaken, 
while that from $jj+\MET$ is slightly tightened 
especially in large $m_X$ region.

We refer to the upper bounds on the number of events 
 in the signal regions defined in the $jj+\MET$ search,  
that require more than two energetic jets and large effective mass, 
defined as a sum of $\MET$ and $p_T$ of jets with $p_T > 50$ GeV. 
Even events with hadronically decaying top quarks can contribute 
to these signal regions 
although the produced jets tend to be softer than the ones in the cases 
that the extra quarks decay to the light quarks directly. 
Since half of the up-type extra quark decays to a charm quark in the case (B), 
the limit is slightly tightened compared with the case (A).

The $bb+\MET$ search is dedicated 
to exotic particles that decay to a bottom quark and a invisible particle exclusively, 
and the definitions for the signal regions require 
that there are 2 b-tagged jets 
and $p_T$ of the fourth jet, ordered in $p_T$, is smaller than 50 GeV 
if it exists in an event. 
These cuts reject the events unless both of the extra quarks decay to a bottom quark. 
This fact indicates 
that the branching fraction of the down-type extra quark 
influences limits from the $bb + \MET$ search significantly. 
We also found that 
the t-channel $X$ exchanging production process induced by the large Yukawa coupling 
is not important 
and the production process is governed by the usual QCD processes.

We also evaluate limits on the extra leptons from the $ll+\MET$ search, 
and the result will be shown in Fig.~\ref{fig;BtoKll2}. 
The extra lepton lighter than 500 GeV has already been excluded 
if the DM is lighter than 300 GeV. 
This bound is almost independent of the Yukawa coupling 
and the extra leptons are produced by the Drell-Yan process.

\section{Flavor physics }
\label{sec4}
Based on the study in Sec. \ref{sec3}, we investigate the
flavor physics especially concerned with the $b \to s ll$ processes,
where the LHCb collaboration has reported the deviations from the SM predictions.
In our model, the $b \to s$ transition is induced by the box diagram involving $Q^\prime$, $\lv$ and $X$.
We notice that the operator relevant to the $B_s$-$\overline{B_s}$ mixing is also generated by
the similar box diagram involving $Q^\prime$ and $X$.
In this section, we discuss not only the rare $B$ decay, but also the other flavor processes relevant to our scenario.

\begin{table}[t]
\begin{center}
  \begin{tabular}{|c|c||c|c|} \hline
 $m_d$(2 GeV) & 4.8$^{+0.5}_{-0.3}$ MeV \cite{PDG}  &  $\lambda$& 0.22509$^{+0.00029}_{-0.00028}$  \cite{CKMfitter}     \\ 
    $m_s$(2 GeV) & 95$\pm 5$ MeV \cite{PDG} & $A$& $0.8250^{+0.0071}_{-0.0111}$  \cite{CKMfitter}  \\ 
      $m_{b}(m_b)$&4.18$\pm 0.03$ GeV  \cite{PDG}    &  $\overline{\rho}$& 0.1598$^{+0.0076}_{-0.0072}$  \cite{CKMfitter} \\ 
  $\frac{2m_{s}}{(m_u + m_d)}$(2 GeV)& 27.5$\pm 1.0$ \cite{PDG}   &  $\overline{\eta}$&   0.3499$^{+0.0063}_{-0.0061}$ \cite{CKMfitter} \\ 
     $m_{c}(m_c)$&1.275$\pm 0.025$ GeV  \cite{PDG}  & $M_Z$ & 91.1876(21) GeV  \cite{PDG}  \\ 
    $m_t(m_t)$& 160$^{+5}_{-4}$ GeV  \cite{PDG}  &  $M_W$ & $80.385(15)$ GeV  \cite{PDG}   \\ 
 $\alpha$ & 1/137.036  \cite{PDG}  &  $G_F$  & 1.1663787(6)$\times 10^{-5}$ GeV$^{-2}$  \cite{PDG}       \\ 
 $\alpha_s(M_Z)$ & $0.1193(16)$ \cite{PDG}   &  &
   \\ \hline
  \end{tabular}
 \caption{The input parameters in our analysis. The CKM matrix, $V$, is written in terms of $\lambda$, $A$, $\overline{\rho}$ and $\overline{\eta}$ \cite{PDG}.}
  \label{table;input}
  \end{center}
\end{table}

\subsection{Constraints from the $B_s$-$\overline{B_s}$ mixing}
In Refs. \cite{Okawa,Okawa2}, the contributions to $\Delta F=2$ processes have been studied in this kind of model.
Now, we are especially interested in the $B_s$-$\overline{B_s}$ mixing, since that process directly relates to
the $b \to s$ transition. 
The box diagram involving $X$ and $Q^\prime$
 induces the operators relevant to the $B_s$-$\overline{B_s}$ mixing:
\beq
{\cal H}^{\Delta F=2}_{eff}= (C_1)_{sb} (\overline{s_L} \gamma^\mu b_L)  (\overline{s_L} \gamma^\mu b_L) +h.c..
\eeq
The Wilson coefficient at the one-loop level is given by \cite{Okawa},
\beq
(C_1)_{sb}=  \frac{( \lambda_{ b}  \lambda^{*}_s  )^2}{64 \pi^2} \frac{1}{(m^{ 2}_{Q^\prime}-m^2_X)^2}
\left \{ \frac{m^{ 2}_{Q^\prime}+m^2_X}{2} + \frac{m^2_X m^{ 2}_{Q^\prime}}{m^{ 2}_{Q^\prime}-m^2_X} \ln\left ( \frac{m^2_X}{m^{2}_{Q^\prime}} \right ) \right \}.
\eeq

One important observable of the $B_s$-$\overline{B_s}$ mixing is the mass difference, $\Delta m_s$, 
that is measured with high accuracy: $\Delta m_s = 17.757\pm 0.021$ ps$^{-1}$~\cite{Amhis:2014hma}.
The SM prediction, however, still suffers from the large uncertainty of the form factor:
$f_{B_s} \hat{B}_{B_s}^{1/2}= 0.266\pm 0.018$~\cite{Aoki:2016frl}, where
$f_{B_s}$ is the decay constant of the $B_s$ meson and $\hat{B}_{B_s}$ is the bag parameter.
We obtain the SM prediction
as $(\Delta m_s)_{{\rm SM}}=18.358$ ps$^{-1}$, using $m_{B_s}= 5.3663$ GeV and $\eta_B=0.55$ \cite{Buras:1990fn}.
The input parameters used in our analysis are summarized in Table \ref{table;input}.
Taking into account the error, the SM prediction is consistent with the experimental result.
Therefore, we estimate the upper limit of the deviation from the SM prediction within $1 \sigma$
and draw the bound on the parameters in this model.

The upper bound on the deviation from the SM prediction is estimated as about 14~\%.
When $|\lambda_b \lambda_s|$ is fixed at 0.15, the lower bound on the DM mass
is estimated as follows:
\beq
\label{eq;BsBsbar}
m_X \geq 593\, {\rm GeV}~ (478\, {\rm GeV}),
\eeq
at $m_{Q^\prime}=1$ TeV ($1.1$ TeV).

\subsection{The excesses of the $b \to s \,  ll$ processes}

\begin{figure}[!t]
\begin{center}
{\epsfig{figure=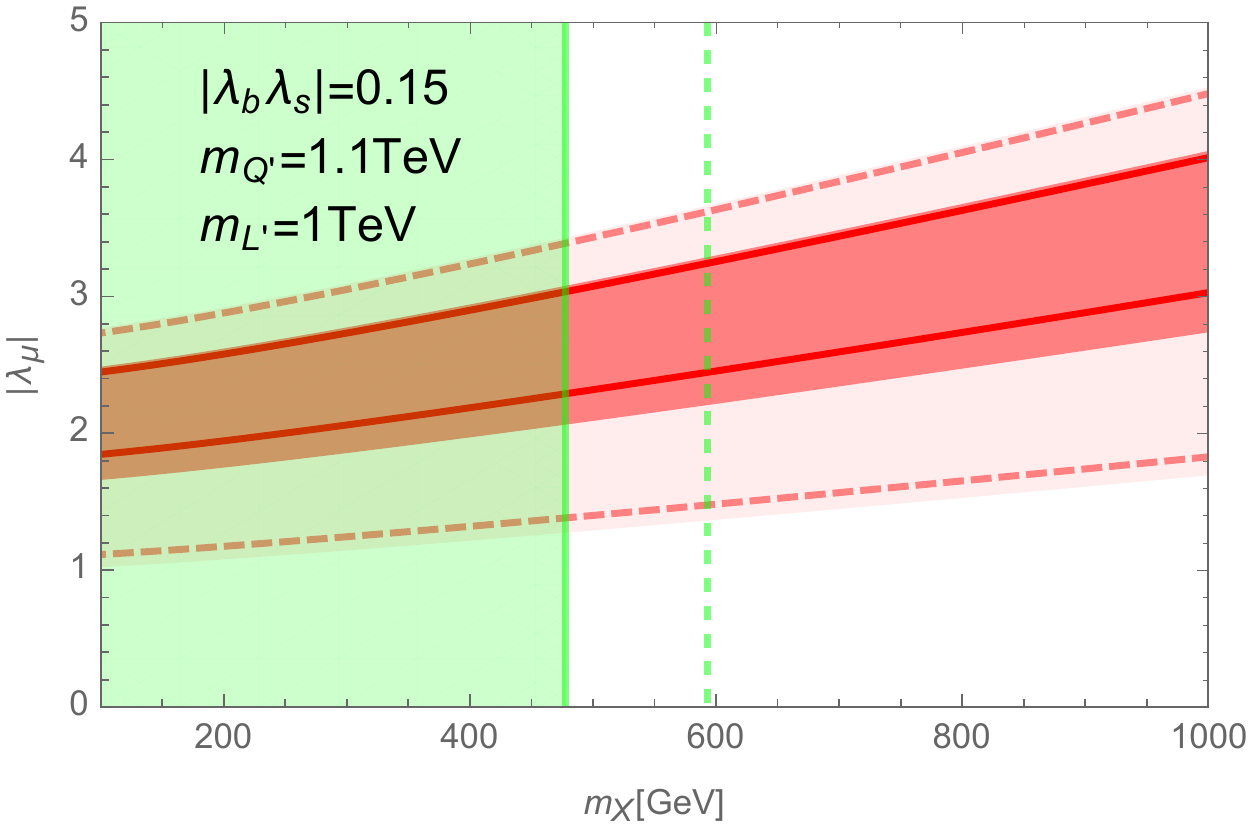,width=0.5\textwidth}} 
\caption{ The required values of $\lambda_\mu$ for the $R_K$ and $R_{K^*}$ anomalies when $m_{\lv}=1$ TeV, $ |\lambda_b \lambda_s|=0.15$ and
$m_{Q^\prime}=1.1$ TeV. The (light) red region is the $1 \sigma$ ($2\sigma$) region of $R_K$. The region for $R_{K^*}$
is not so different from the $R_K$. The bands within thick and dashed red lines correspond to  $1 \sigma$ and $2\sigma$ regions of $R_{K^*}$. The (dashed) green line depicts the lower limit
from $B_s$-$\overline{B_s}$ mixing in Eq. (\ref{eq;BsBsbar}) at $m_{Q^\prime}=1.1$ TeV (1 TeV).}
\label{fig;BtoKll}
\end{center}
\end{figure}

Finally, we consider the $b \to s l l$ ($l=e, \, \mu$) decays. 
The explanation of the anomaly has been done in the setups similar to our model \cite{Gripaios:2015gra,Arnan:2016cpy}.
In our setup, the box diagram, that involves $X$, $Q^\prime$ and $\lv$, contributes to the flavor violating processes, 
$b\to s ll$.  The $\Delta B = 1$ effective Hamiltonian is given by
\begin{equation}
{\cal H}_\textrm{eff}= -g_{\rm SM}
\left  \{ C_9^{l}(\overline{s_L} \gamma_\mu b_L)
(\overline{l}\gamma^\mu l )
+C_{10}^{l}(\overline{s_L} \gamma_\mu b_L)
(\overline{l}\gamma^\mu \gamma_5 l) +h.c. \right \},
\end{equation}
where $g_{\rm SM}$ is the factor from the SM contribution:
\begin{equation}
g_{\rm SM} =\frac{4 G_F}{\sqrt{2}} V_{tb} V^*_{ts} \frac{e^2}{16 \pi^2}.
\end{equation}
The Wilson coefficients $C_{9}^{l}$ and $C_{10}^{l}$ 
consist of the SM and the new physics contributions
as $C_{9}^{l}=(C_{9}^{l})_{{\rm SM}} +\Delta C_{9}^{l}$ and $C_{10}^{l}=(C_{10}^{l})_{{\rm SM}} +\Delta C_{10}^{l}$.
The new physics contributions in this model are given by
\begin{equation}
\Delta C_{9}^{\mu} =- \Delta C_{10}^{\mu}  = \frac{\lambda_{ b}  \lambda^{*}_s |\lambda_\mu|^2}{128 \pi^2 g_{\rm SM}} \frac{1}{m^{ 2}_{Q^\prime}-m^2_{L^\prime}}
\left \{ f(m^2_X/m^{ 2}_{Q^\prime})-f(m^2_X/m^{ 2}_{L^\prime}) \right \},
\end{equation}
where $f(x)$ is defined as
\beq
f(x)=\frac{1}{x-1}-\frac{\ln x}{(x-1)^2}.
\eeq

The branching fractions BR($B^+ \to K^+ \, ll$) and BR($B_0 \to K^* \,ll$) are recently measured in the LHCb experiment \cite{Aaij:2014ora,LHCbnew}. 
In the SM, the ratio between the branching fractions for $l=\mu$ and $l=e$  
is predicted to be almost unity. 
The measured values of such observables, namely $R_K$ and $R_{K^*}$, are reported 
in each bin of $q^2$ GeV$^2$, which is the invariant mass of two leptons in the final state \cite{Aaij:2014ora,LHCbnew}. 
In particular, both results in $B^+ \to K^+ \, \mu \mu$ and $B_0 \to K^* \, \mu \mu$ with $1$ GeV$^2$ $\leq q^2 \leq 6$ GeV$^2$ 
are smaller than the SM predictions, 
and the deviations of $R_K$ and $R_{K^*}$ exceed $2 \sigma$: $R_K=0.745 \pm 0.097$ \cite{Aaij:2014ora} and $R_{K^*}=0.685 \pm 0.122$ \cite{LHCbnew}. In the region with lower $q^2$, we can also see the deviation of $R_{K^*}$, although
the contribution from the electric dipole operator needs to be taken into account \cite{LHCbnew}.

As discussed in Refs. \cite{Descotes-Genon:2013wba,Hiller:2014yaa,Altmannshofer:2015sma,Descotes-Genon:2015uva,Hurth:2016fbr,Altmannshofer:2017yso,DAmico:2017mtc,Geng:2017svp,Ciuchini:2017mik,Alok:2017jaf},
the new physics contributions to $C_{9}^{l}$ and $C_{10}^{l}$ are required to explain the excesses.
$R_K$ and $R_{K^*}$ including the new physics contributions, for instance, are calculated in Refs. \cite{Descotes-Genon:2015uva,Celis:2017doq,Bardhan:2017xcc}.

Figure~\ref{fig;BtoKll} shows the required values of $\lambda_\mu$ for the $R_K$ and $R_{K^*}$ anomalies when $m_{L^\prime}=1$~TeV, $ |\lambda_b \lambda_s|=0.15$ and $m_{Q^\prime}=1.1$~TeV. 
The (light) red region is the $1 \sigma$ ($2\sigma$) region of $R_K$. The bands within thick and dashed red lines correspond to  $1 \sigma$ and $2\sigma$ regions of $R_{K^*}$. The (dashed) green line depicts the lower limit
from the $B_s$-$\overline{B_s}$ mixing in Eq. (\ref{eq;BsBsbar}) at $m_{Q^\prime}=1.1$ TeV (1 TeV).
We see that $\lambda_\mu$ need to be about $2$ to explain the excesses within $1 \sigma$.

\begin{figure}[!t]
\begin{center}
{\epsfig{figure=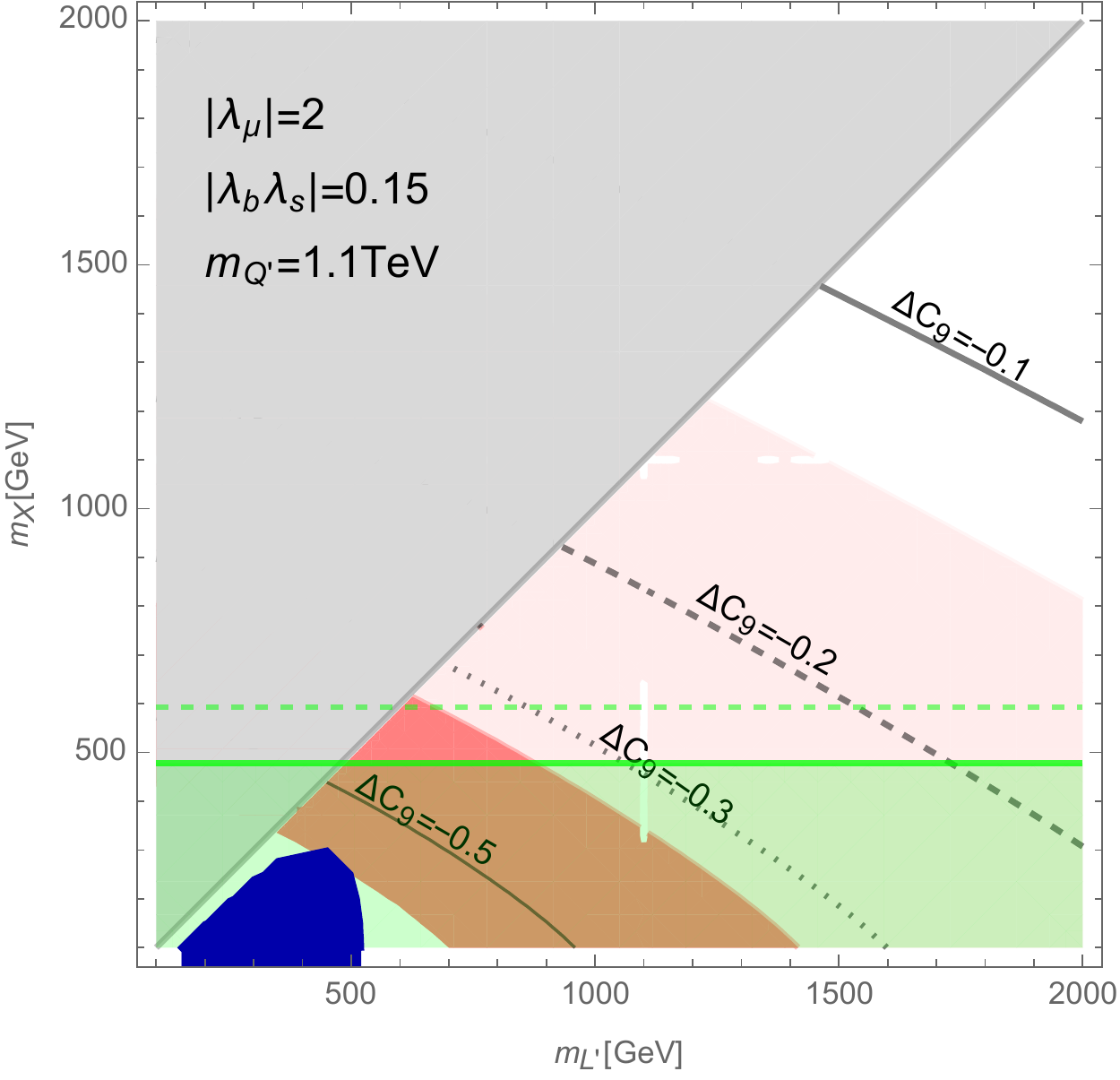,width=0.5\textwidth}} 
\caption{ $\Delta C_9$ on the plane of $m_{L^\prime}$ and $m_X$ with $ |\lambda_\mu|=2$, $ |\lambda_b \lambda_s|=0.15$ and $m_{Q^\prime}=1.1$~TeV. The size of $\Delta C_9=-\Delta C_{10}$ on each black line
is $-0.1$ (thick), $-0.2$ (dashed), $-0.3$ (dotted), and $-0.5$ (solid), respectively.
The (light) red region is the $1 \sigma$ ($2\sigma$) region of $R_K$. The (dashed) green line depicts the lower limit
from $B_s-\overline{B_s}$ mixing in Eq. (\ref{eq;BsBsbar}) at $m_{Q^\prime}=1.1$~TeV (1~TeV). The blue region is excluded by $\mu \mu + \MET$ at the LHC \cite{llMET}.}
\label{fig;BtoKll2}
\end{center}
\end{figure}

In Fig. \ref{fig;BtoKll2}, the predicted $\Delta C_{9}^{l}$ is also depicted on the $m_{L^\prime}$-$m_X$ plane,
when $\lambda_\mu$ is fixed at $\lambda_\mu=2$.
The value required by the global fitting, where the angular distribution of $B \to K^* \mu \mu$ is also included, is about $25$ \%, compared to the SM prediction.
The magnitudes of $(C_{9}^{l})_{{\rm SM}}$ and $(C_{10}^{l})_{{\rm SM}}$ are about $4$, so that
${\cal O}(1)$ value of $|\Delta C_{9}^{\mu}|$ seems to be required by the global analysis.
As we see in Fig. \ref{fig;BtoKll2}, the region with $|\Delta C_{9}^{l}| \gtrsim 0.5$ is below the
exclusion limit from the $B_s$-$\overline{B_s}$ mixing. 
The $1 \sigma$ region of $|\Delta C_{9}^{\mu}|$ suggested by the global analysis, for instance,
is $-0.81 \leq \Delta C_{9}^{\mu} \leq -0.48$ ($1 \sigma$) and $-1.00 \leq \Delta C_{9}^{\mu} \leq -0.32$ ($2 \sigma$) \cite{Altmannshofer:2017yso}. Then, we could conclude that
our model can fit the experimental results including the angular distribution of $B \to K^* \mu \mu$
within $2 \sigma$ but not $1 \sigma$ as far as $\lambda_\mu$ is equal to $2$.

In the next section, we study DM physics based on these analyses.
In this kind of model, large Yukawa couplings are required by the thermal relic density,
because the DM candidate is a complex scalar. Then, we will find the parameter region
that the explanation of the excesses within $2 \sigma$ as well as the DM relic density can be achieved.

Before the DM physics, let us discuss the other observables in flavor physics.
We simply assume that $\lambda_d$ is tiny to avoid the strong constraints from $K_0$-$\overline{K_0}$
and $B_d$-$\overline{B_d}$ mixings.
In this setup, since SU(2)$_{\rm L}$ doublet quarks couple to the extra quarks and DM, 
the sizable $\lambda_b$ and $\lambda_s$ generate sizable
Yukawa couplings between left-handed up-type quarks and the extra up-type quarks as well.
Then, the constraint from $D_0$-$\overline{D_0}$ should be taken into account.
The relevant Yukawa couplings are as follows:
\beq
\label{Yukawa-D}
(\lambda_{ s} V^*_{us} +\lambda_{ b} V^*_{ub} )\overline{Q^\prime_{1 \, R}}  X^\dagger u_{L}  +(\lambda_{ s} V^*_{cs} +\lambda_{ b} V^*_{cb} )\overline{Q^\prime_{1 \, R}}  X^\dagger c_{L}. 
\eeq
Here, $Q'_1$ denotes the isospin 1/2 component of $Q'$. 
Each coupling is suppressed by the CKM matrix, 
and the most relevant coupling is expected to be induced by 
$\lambda_{ s} $ because of the relatively large elements of the CKM matrix.

On the other hand, the theoretical prediction of $D_0$-$\overline{D_0}$ also suffers from 
the ambiguity of the long-distance correction \cite{Golowich:2007ka,Petrov:2013usa}.
Therefore, we estimate the new physics contributions and compare the order of the SM prediction.
The Yukawa couplings in Eq. (\ref{Yukawa-D}) induce the $\Delta F=2$ operator with $(C_1)_{uc}$
via the box diagram. The observable of the $D_0$-$\overline{D_0}$ mixing is given by
\beq
x_D= 2 |M^D_{12}| \, \tau_D,
\eeq
where $\tau_D$ is the life time: $\tau_D=0.41$ ps \cite{PDG}.
$M^D_{12}$ includes both the SM prediction and the new physics contribution.
Using the input parameters in Refs. \cite{PDG, Carrasco:2014uya,Amhis:2016xyh}, the new physics contribution to $x_D$ is estimated as $\sim 2 \times 10^{-4}$ at $(\lambda_b,\, \lambda_s)=(1,\,0.15)$, $m_{Q^\prime}=1$~TeV and $m_X=500$~GeV.
The SM prediction is ${\cal O}(10^{-2})$ \cite{Petrov:2013usa}, so that these parameters are safe for the $D_0$-$\overline{D_0}$ mixing.
The upper bound on $\lambda_s$ would be about 0.3. 
If $\lambda_s$ is ${\cal O}(1)$ and $\lambda_b$ is ${\cal O}(0.1)$,
the new physics contribution becomes hundreds times bigger than the SM prediction.
Thus, we concentrate on the case (A) and the case (B) 
where $|\lambda_b| \geq |\lambda_s|$ is satisfied.

\section{The impact on DM physics}
\label{sec5}

Here, we study DM physics and discuss the consistency between the explanation of the $b \to s ll$ anomalies and the DM observations. 
As discussed above, the $b \to s ll$ anomalies and the constraints from the $\Delta{F}=2$ processes 
imply that $|\lambda_b \lambda_s|=0.15$ and $|\lambda_b| \geq |\lambda_s|$ for $m_{Q'} =$ 1.1 TeV.  
Then, in this section, we analyze the DM relic density and the direct detection bounds, focusing on two parameter sets: 
$(\lambda_b, \,\lambda_s)=(1.0, \, 0.15)$ (Case A) and $(\lambda_b, \,\lambda_s)=(\sqrt{0.15}, \,\sqrt{0.15})$ (Case B). In both cases, $m_{Q'}$ is fixed at 1.1 TeV. 
Note that we employ micrOMEGAs\_4.3.4~\cite{Belanger:2014vza} to evaluate the DM relic density in our analysis.

\subsection{Relic density}

\begin{figure}[!t]
\begin{center}
{\epsfig{figure=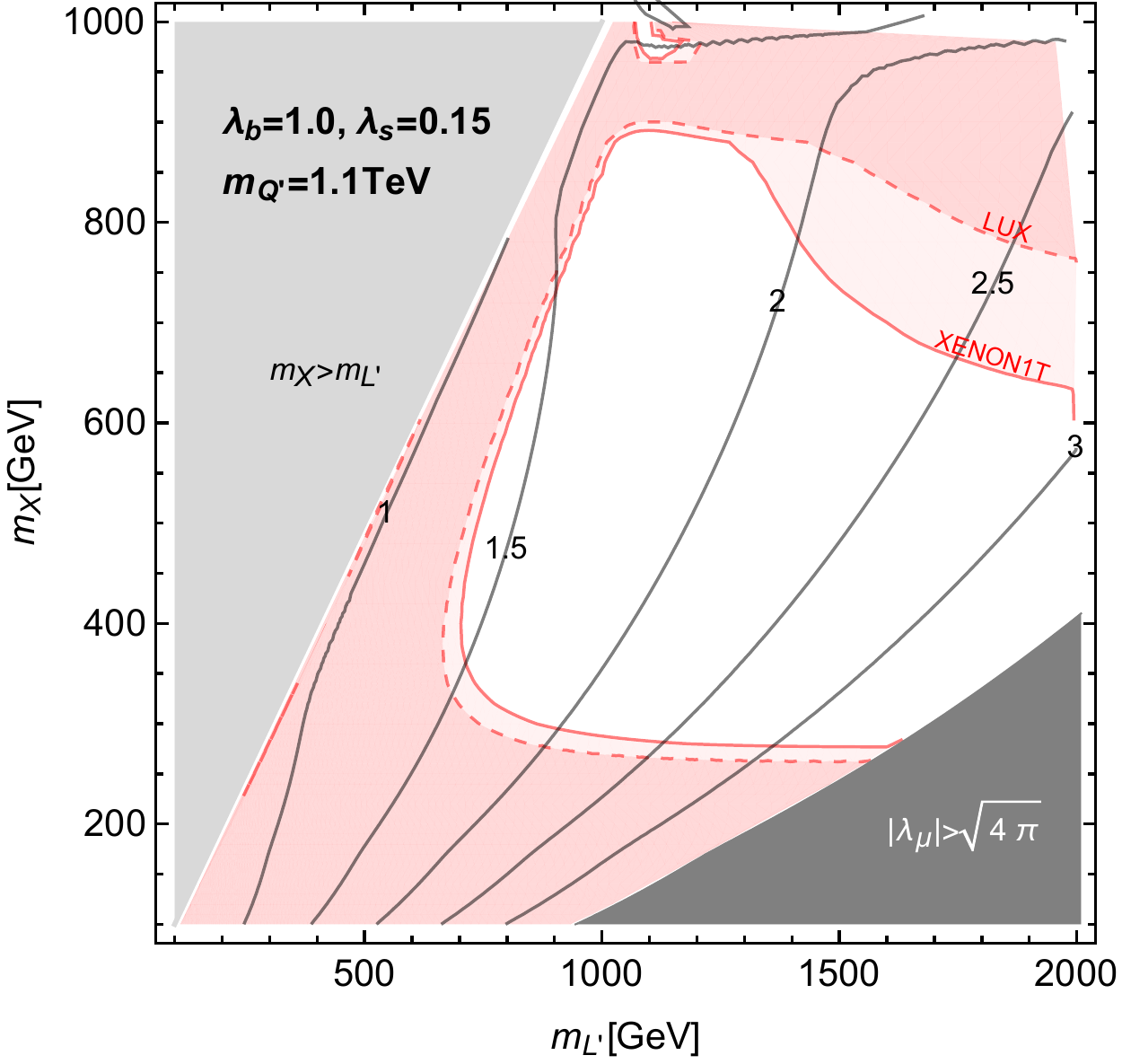,width=0.45\textwidth}} 
{\epsfig{figure=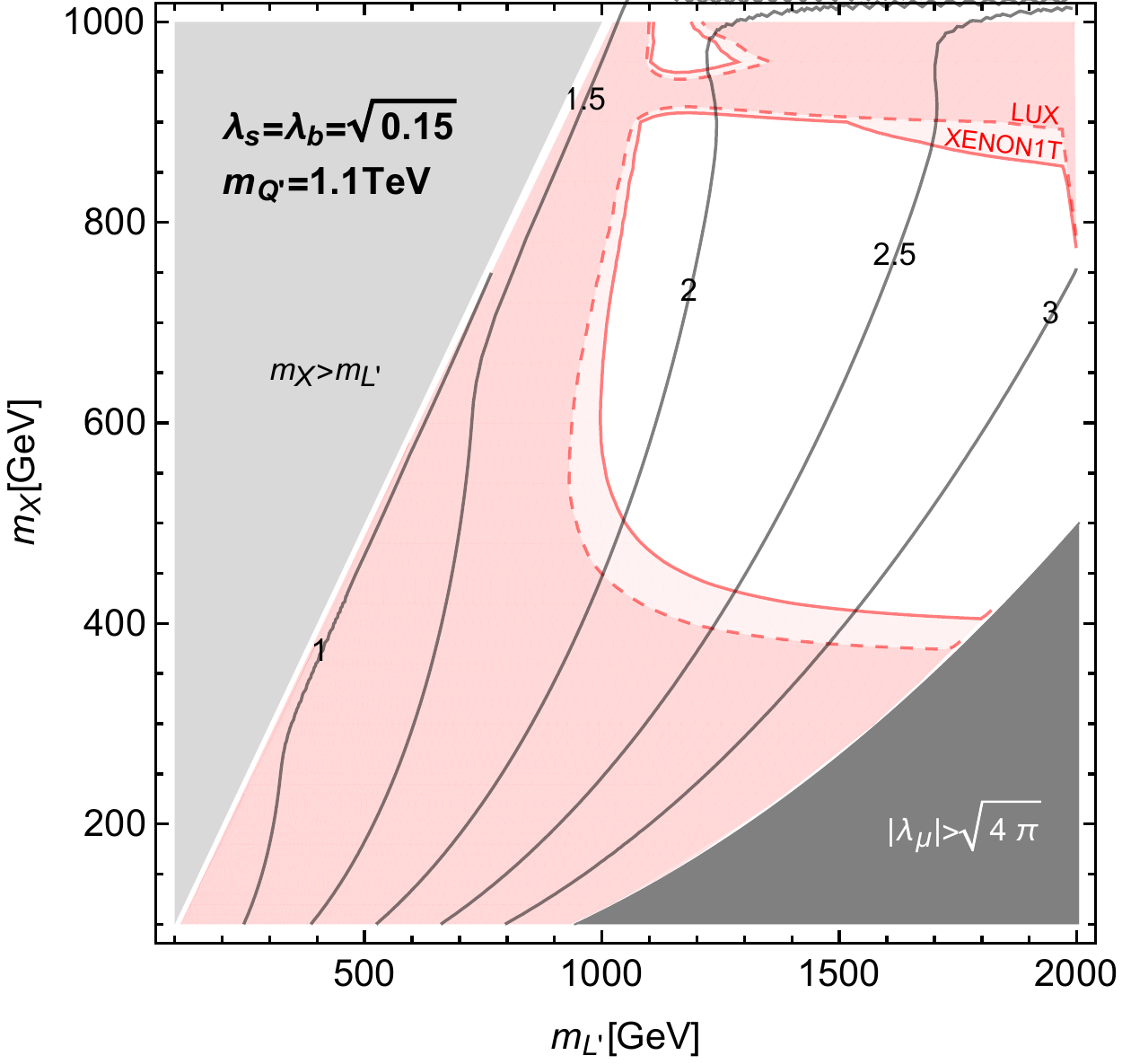,width=0.45\textwidth}} 
\caption{The values of $|\lambda_\mu|$ required for the central value of the observed DM abundance (black solid lines); $\Omega_{\rm CDM} h^2 = 0.1186\pm0.0020$~\cite{PDG}. 
The left (right) panel corresponds to the case A (case B). DM direct detection experiments constrain the red region.}
\label{fig;mXmL_lammu}
\end{center}
\end{figure}
The DM annihilation process is governed by the extra quark and lepton exchanging in $t$-channel: $XX^\dagger \to Q\overline{Q}, l\overline{l}$. 
Since the Yukawa terms in Eq.(\ref{eq;VB}) preserve the chiralities of the interacting fermions, the final state fermions are left-handed in the massless limit. 
Then, the $s$-wave contribution is suppressed by the fermion mass, so that the annihilation becomes $p$-wave dominant. 
That requires the Yukawa coupling to be so large that the observed value of the DM abundance is explained thermally.  
As seen in Fig.~\ref{fig;BtoKll}, the $b \rightarrow s ll$ anomalies also require a large $|\lambda_\mu|$, 
and this fact indicates that  the $b \to s ll$ anomalies can be compatible 
with the observed DM density in our model. 

In Fig.~\ref{fig;mXmL_lammu}, we show the values of $|\lambda_\mu|$, by black lines, required to explain the central value of the observed DM abundance: $\Omega_{\rm CDM} h^2 = 0.1186\pm 0.0020$~\cite{PDG}. 
In the case (B) with $|\lambda_b|=|\lambda_s|=\sqrt{0.15}$, the dominant annihilation process is $XX^\dagger \to \mu \overline{\mu}, \nu \overline{\nu}$ 
in our parameter space, except for the coannihilation region. 
We find in the right panel of Fig.~\ref{fig;mXmL_lammu} 
that $|\lambda_\mu| \simeq 2$ is predicted by $(m_X, m_{\lv})=$(500 GeV, 1 TeV). 
The $b \to sll$ anomalies are simultaneously explained with the DM density 
around this region.    

In the case (A) with $|\lambda_b|=1.0$ and $|\lambda_s|=0.15$, other processes, especially $XX^\dagger \to b\overline{b}, t\overline{t}$, may assist in 
reducing the DM relic density in addition to $XX^\dagger \to \mu \overline{\mu}, \nu \overline{\nu}$. 
Since we fix $m_{Q'}$ at 1.1 TeV, these processes dominate the annihilation process in the region with $m_{\lv} > 1$ TeV. 
In fact, we find in the left panel of Fig.~\ref{fig;mXmL_lammu} that 
the observed density is explained by smaller values of $|\lambda_\mu|$ than the case (B).

\subsection{Direct detection}

\begin{figure}[!t]
\begin{center}
{\epsfig{figure=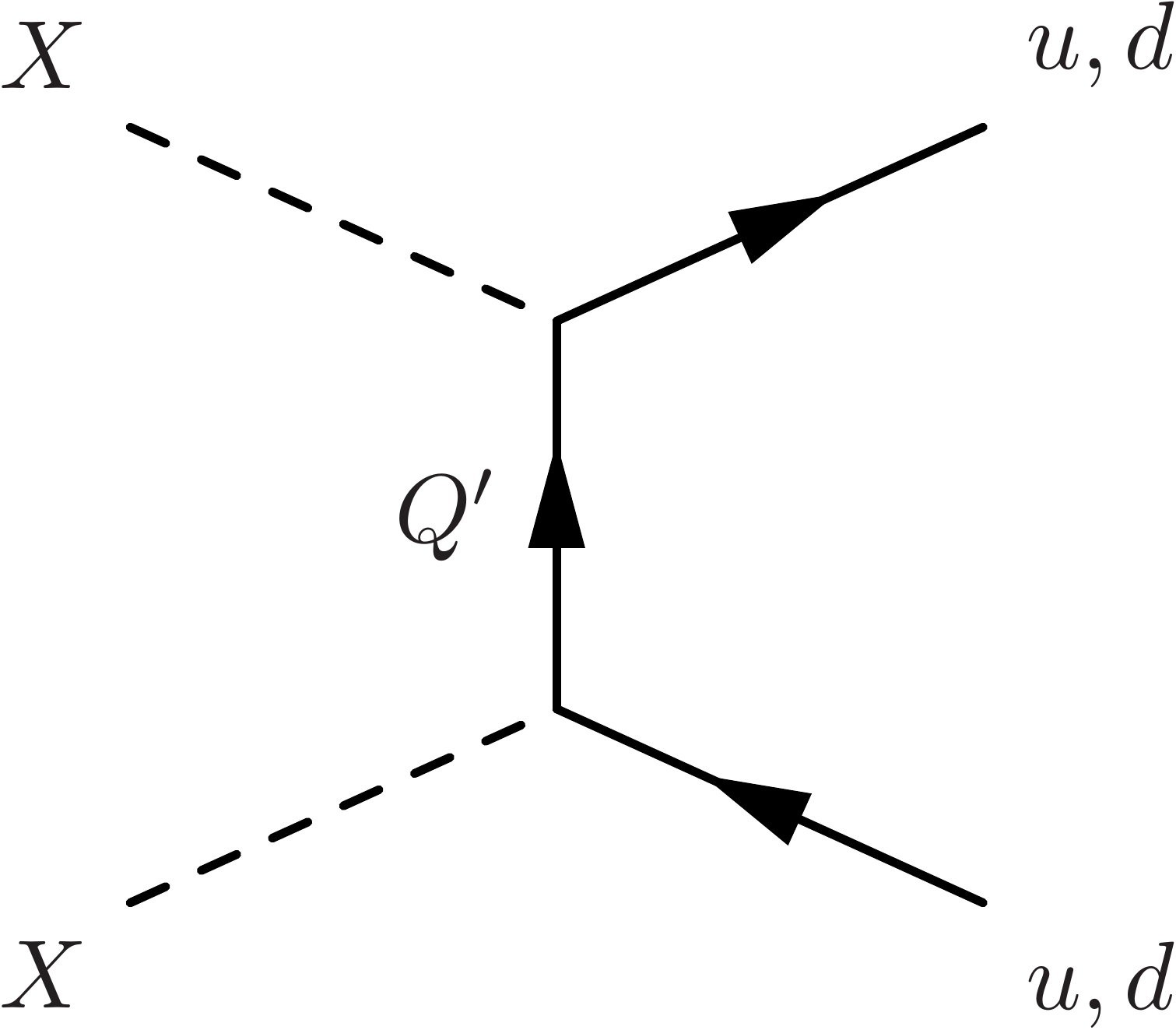,width=0.21\textwidth}}
\hspace{0.5cm}
{\epsfig{figure=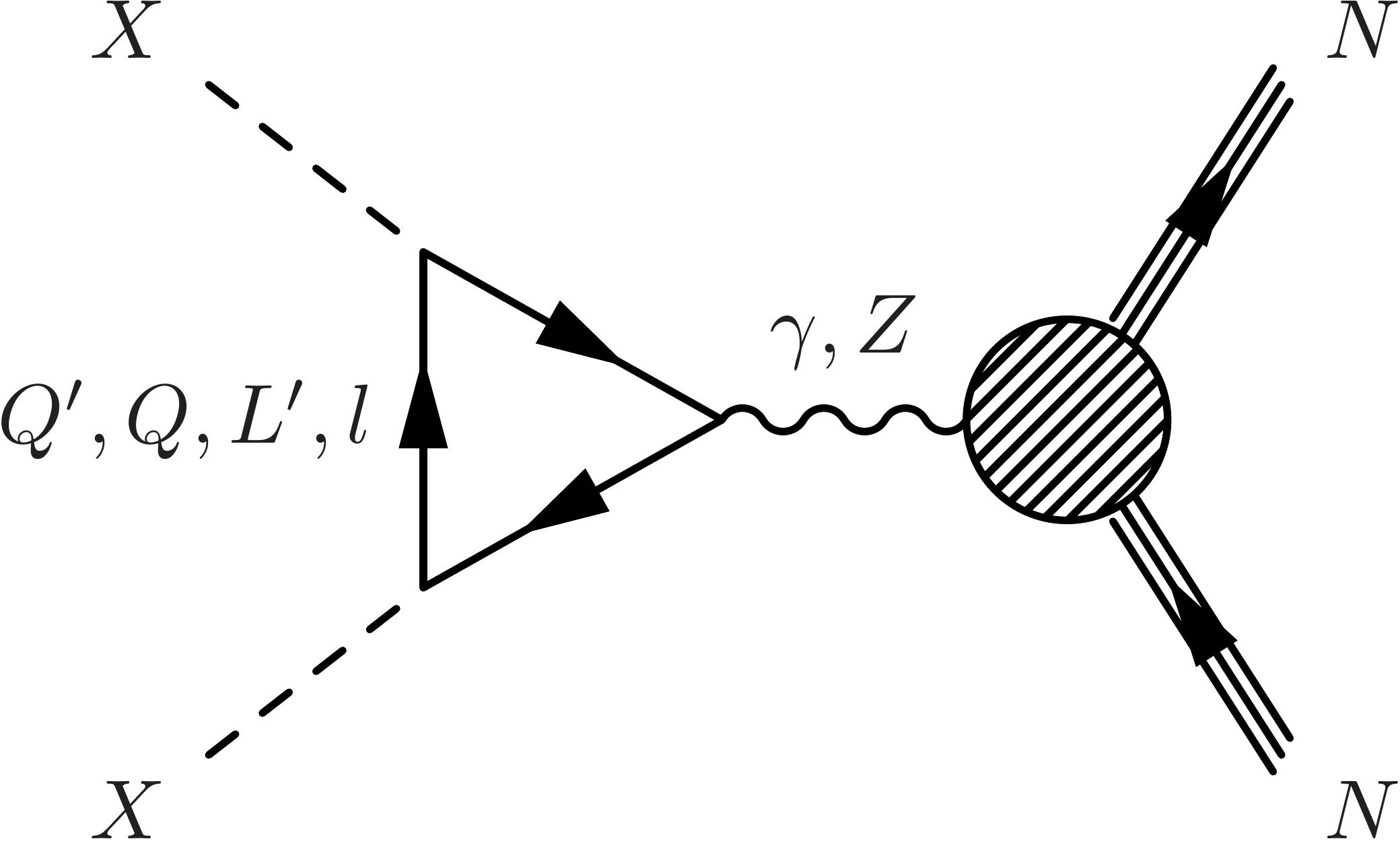,width=0.3\textwidth}} 
\hspace{0.5cm}
{\epsfig{figure=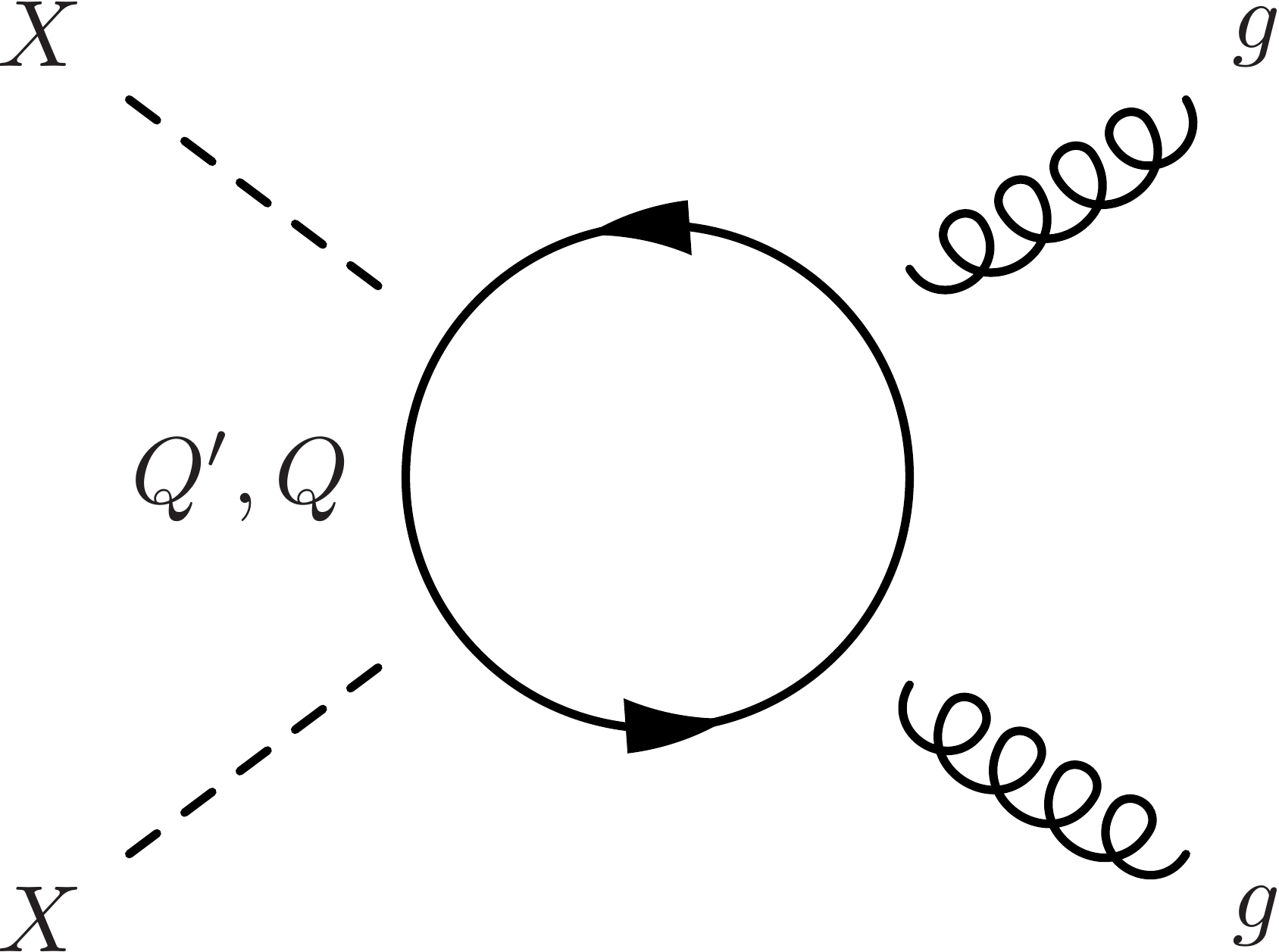,width=0.25\textwidth}}
\caption{Diagrams contributing to the DM-nucleon elastic scattering.}
\label{fig;DD}
\end{center}
\end{figure}
Next, we discuss the constraint from the DM direct detection.
Figure~\ref{fig;DD} shows relevant processes to the DM-nucleon scattering in this model.  
The type of dominant process depends on mass spectrum of the DM 
and the extra fermion and the size of the Yukawa couplings. 

If the DM has a sizable interaction with the up or down quarks, 
the dominant contribution to the direct detection will be the tree-level scattering 
through the extra quark exchanging, as the left diagram in Fig.~\ref{fig;DD}. 
This process is not suppressed by nucleon form factors, and thus leads a large cross section.   
Note that in our model, even if $\lambda_d$ is vanishing, the DM can interact with the up quark through the CKM matrix as in Eq.(\ref{Yukawa-D}). 

There are large contributions from the 1-loop vector-boson exchanging, as shown in the center of Fig.~\ref{fig;DD}, because the large Yukawa coupling is required by the DM relic density. 
As studied in the literatures~\cite{Agrawal:2011ze,Bhattacharya:2015xha,Okawa}, the triangle diagram involving the SM fermion $f$ and the extra fermion $F$ induces 
the scattering of DM with the proton through the photon exchanging, and the contribution is given by 
\beq
\label{photon}
f_p^{\rm photon} \simeq \frac{ e^2 Q_f |\lambda_f|^2}{16\pi^2 m_F^2} \frac{N_c}{3} \ln\left(\frac{m_f^2}{m_F^2}\right) ,
\eeq
in the limit that $m_F \gg m_X, m_f$. 
$N_c$ and $Q_f$ denote the number of color 
and the electromagnetic charge for the fermion $f$. 
It is interesting that there is a logarithm enhancement as $\ln(m_f^2/m_F^2)$, so that the lighter fermion has larger contribution. 

In addition, the diagram similar to the photon exchanging also generates the $Z$-boson exchanging process. 
The contribution is evaluated as 
\beq
\label{Zexchange}
f_p^{Z} = (4s_W^2-1) \frac{G_F a_Z}{\sqrt{2}} ,\quad
f_n^{Z} = \frac{G_F a_Z}{\sqrt{2}} ,
\eeq
where $s_W$ denotes the sine of the Weinberg angle and $a_Z$ is a loop function:
\beq
\label{Zloop}
a_Z \simeq \frac{T_3^f N_c |\lambda_f|^2}{16\pi^2} \frac{m_f^2}{m_F^2} \left[ \frac{3}{2} + \ln\left(\frac{m_f^2}{m_F^2}\right) \right] , 
\eeq
with $T_3^f$ being isospin of the SM fermion $f$ in the loop. 
Since the $Z$-exchanging contribution is proportional to $m_f^2/m_F^2$, 
this process is significant in the case that $f$ is the top quark. 
In our analysis, we use exact expressions for these vector-boson exchanging contributions, not the approximate forms, Eqs.(\ref{photon}) and (\ref{Zloop}). 

Besides, DM can also scatter off the gluon in the nucleon through the box diagram involving the heavy and extra quarks, depicted as the right piece of Fig.~\ref{fig;DD}. 
This contribution is less than 10\% of the photon exchanging contribution 
in most of our parameter space, as discussed in Ref~\cite{Okawa}. 
However, this process is largely enhanced to be dominant contribution 
in the region of $m_{Q'} - m_X \simeq m_q$~\cite{Drees:1993bu,Hisano:2011um,Hisano:2015bma}. 
This contribution is included in our analysis by using micrOMEGAs. 

In the case (B) with $|\lambda_b|=|\lambda_s|=\sqrt{0.15}$, 
the DM significantly interacts with the leptons. 
The Yukawa interaction involving the lepton does not induce the tree-level process, 
and the dominant process is the 1-loop photon exchanging 
in most of our parameter space. 
On the other hand, $\lambda_s$ induces the sizable tree-level scattering with the up quark via the CKM mixing. 
Although $\lambda_s$ is much smaller than $\lambda_\mu$, 
it is possible that $\lambda_s$ gives sizable contributions to the direct detection. 
This process becomes dominant in the region where both $m_X$ and $m_{\lv}$ are large
even when we fix $m_{Q'}$ at 1.1~TeV and much larger than $m_{\lv}$. 
In the right panel of Fig.~\ref{fig;mXmL_C9}, we find that 400 GeV $\lesssim m_X \lesssim$ 900 GeV and $m_{L'} \gtrsim 1$ TeV are allowed by the XENON1T experiment~\cite{XENON1T}. 

In the case (A) of $|\lambda_b|=1$ and $|\lambda_s|=0.15$, 
there is a large contribution from the $Z$-boson exchanging via the top-quark loop, 
in addition to the photon exchanging. 
Even in this case, the photon exchanging is dominant process in the smaller mass region. 
The $Z$-boson exchanging dominates the DM-nucleon scattering in $m_X \gtrsim 400$ GeV and $m_{\lv} \gtrsim 1$ TeV, except for the region 
where $m_{Q'} - m_X \simeq m_t$ and the gluon scattering is dominant. 
Moreover, 
we emphasize that the photon and $Z$-boson exchanging are comparable 
and the sign is opposite. 
Thus there is a considerable cancellation between them. 
In the left panel of Fig.~\ref{fig;mXmL_C9}, we find that the allowed DM mass is 300 GeV $\lesssim m_X \lesssim$ 900 GeV to evade the XENON1T bound~\cite{XENON1T}.

\subsection{$b \to s ll$ anomalies with the observed DM abundance}

\begin{figure}[!t]
\begin{center}
{\epsfig{figure=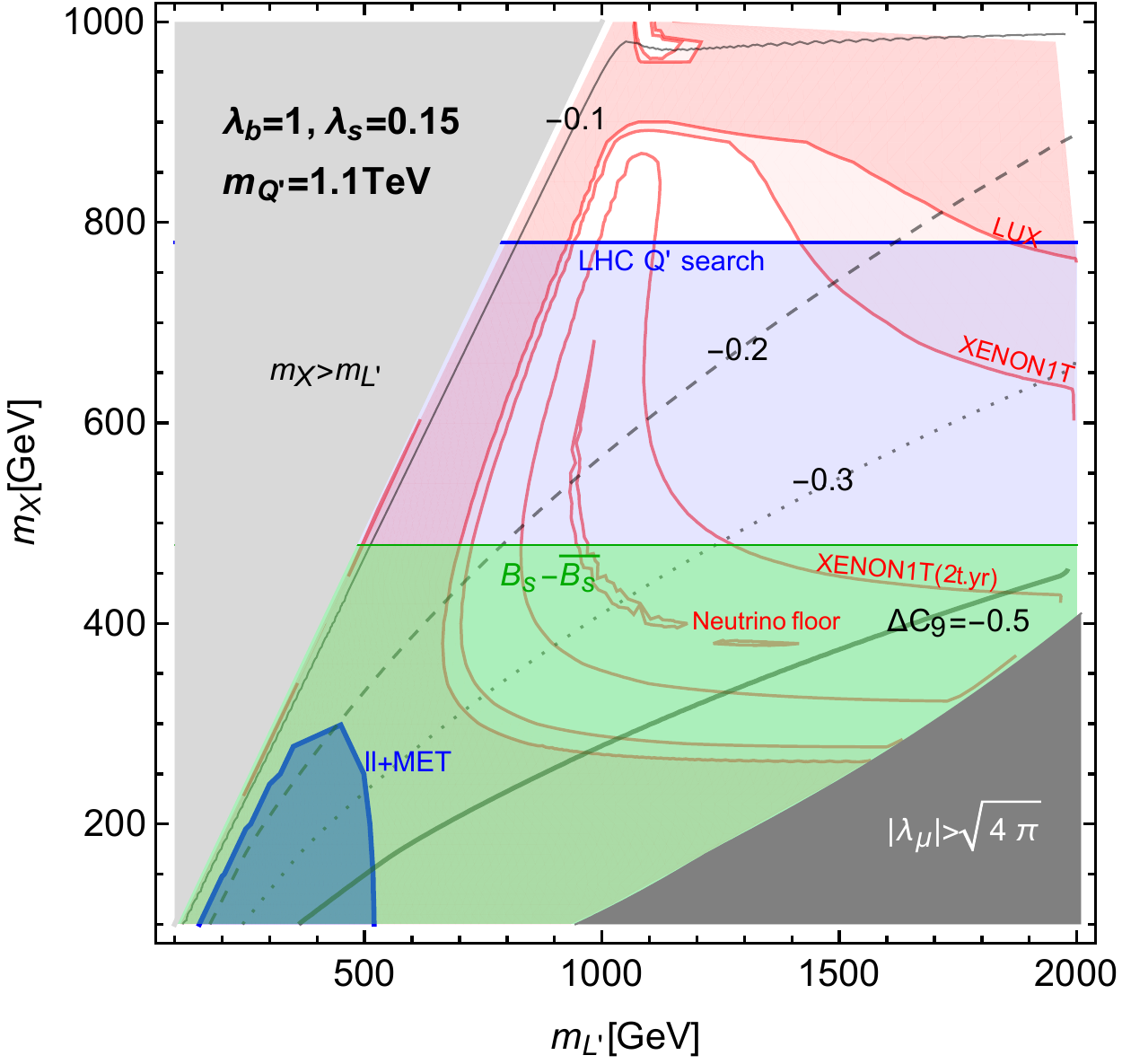,width=0.49\textwidth}}
{\epsfig{figure=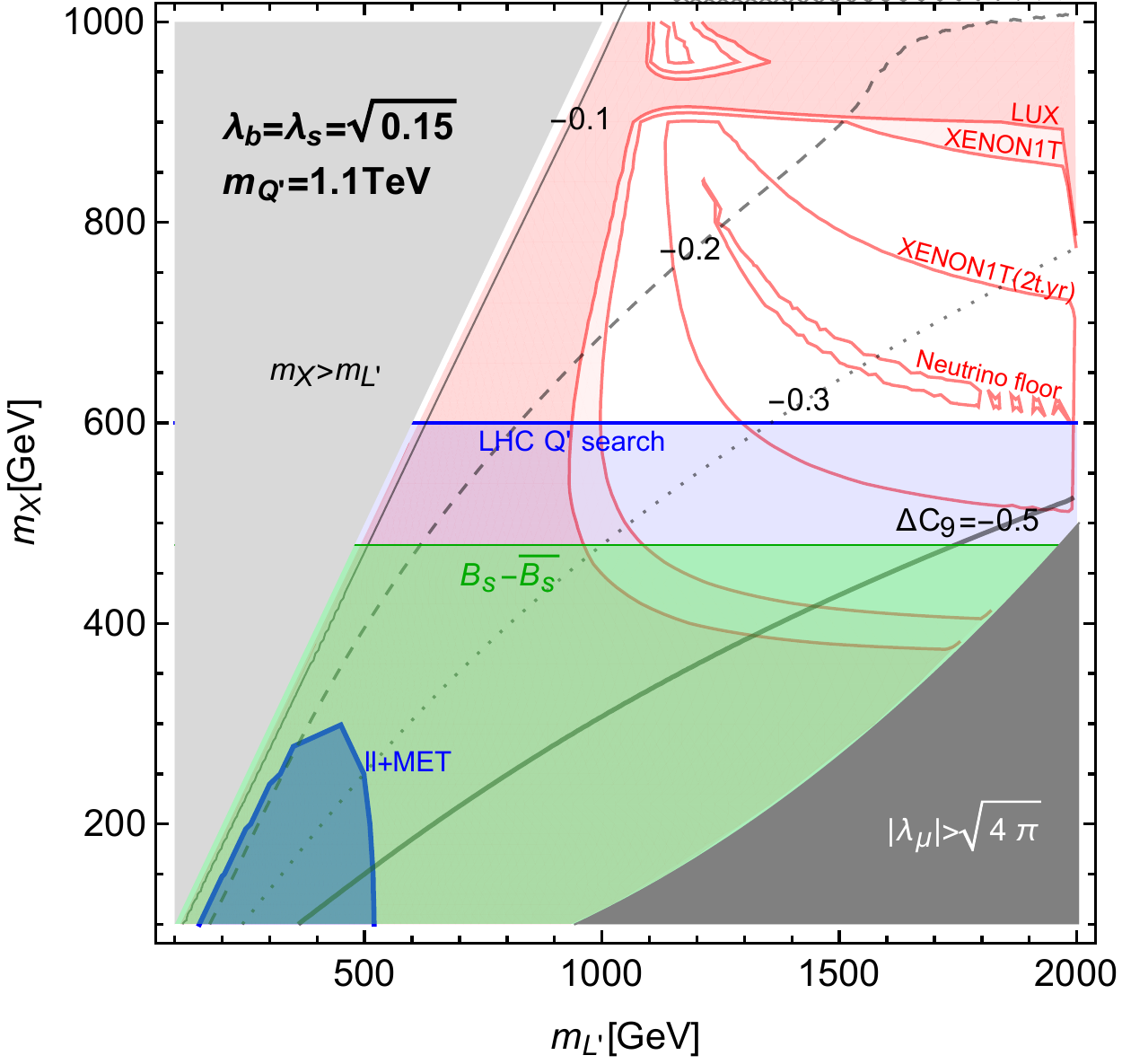,width=0.49\textwidth}}
\caption{The $\Delta C_9^{\mu}$ value (black lines), when the value of $|\lambda_\mu|$ is aligned to explain the observed DM abundance. 
The left (right) panel corresponds to the case A (case B). 
We fill in the region excluded by the LHC, flavor and DM experiments with colors; the extra quark and lepton searches (blue), $B_s$-$\overline{B_s}$ mixing (green) and DM direct detection experiments (red). 
The gray region stands for the perturbativity limit; $|\lambda_\mu| > \sqrt{4\pi}$.  }
\label{fig;mXmL_C9}
\end{center}
\end{figure}

In the Fig.~\ref{fig;mXmL_C9}, we depict the values of $\Delta{C}_9$ by black lines, when $|\lambda_\mu|$ is aligned to account for the measured DM density. 
The left (right) panel corresponds to the case (A) (case (B) ). 
Shaded region is excluded by the LHC, flavor and DM experiments; the extra quark and lepton searches at the LHC (blue), 
the $B_s$-$\overline{B_s}$ mixing observables (green) and the DM direct detections (red). 
We also fill in the region constrained by the theoretical requirement with colors; 
the perturbative bound $|\lambda_\mu| > \sqrt{4\pi}$ (gray) and the unstable $X$, $m_X > m_{L'}$ (light gray). 
Further, we show the future sensitivity prospects in XENON1T experiment and neutrino floor in the figure. 

In the case (A) (left panel), we see that the $Q'$ searches at the LHC and the DM direct detection stringently limit our model parameter. 
Then, the region where $\Delta C_9^\mu \lesssim -0.2$ is excluded by these results.
Following Ref. \cite{Altmannshofer:2017yso}, the global fitting of $\Delta C^\mu_9=-\Delta C^\mu_{10}$ 
suggests $-0.81 \leq \Delta C_{9}^{\mu} \leq -0.48$ ($1 \sigma$) and $-1.00 \leq \Delta C_{9}^{\mu} \leq -0.32$ ($2 \sigma$). Thus, we conclude that it is difficult to explain the $b \to sll$ anomalies within even $2\sigma$ in the case (A). 

In the case (B), on the other hand, we find that even the region where $\Delta C_9^\mu \lesssim -0.3$ is allowed and the $b \to sll$ anomalies can be explained within $2\sigma$, consistently with the other experiments. 

\section{Summary}
\label{sec7}
The LHCb collaboration has reported several excesses in the $b \to s ll$ processes.
A common feature is that the branching ratio of the $b$ decay associated with two muons
is rather small, compared to the SM prediction. 
If the excesses are evidences for new physics, 
they may indicate existences of flavor-violating couplings between the SM fermions and new particles~\footnote{Note that the explanation of the excess without new flavor-violating couplings is also possible \cite{Kamenik:2017tnu}.  }.
So far, many candidates for the new physics have 
been proposed: $Z^\prime$ \cite{Zprime}, leptoquark \cite{leptoquark},  and so on.
In those models, the tree-level diagram involving the extra particle contributes to $C_9$ and $C_{10}$ operators.
Although we may have to introduce many extra fields especially in $Z^\prime$ models to achieve
the realistic Yukawa couplings and the anomaly-free conditions, we can simply explain the excesses without
conflicts with the other flavor observables \cite{Zprime, leptoquark}.

In this paper, we focus on the other simple scenario that extra fields charged under the SM gauge symmetries are introduced and the one-loop diagram involving the extra fields contributes to the $C_9$ and $C_{10}$ operators
\cite{DAmico:2017mtc,Gripaios:2015gra,Arnan:2016cpy,Belanger:2015nma,Hu:2016gpe,Poh:2017tfo,Kamenik:2017tnu}.
In order to explain the experimental results of $R_K^{(*)}$, 
we need rather large Yukawa couplings between 
the extra lepton and muon. The Yukawa couplings in the quark sector, on the other hand,
need to be rather small to avoid the constraint from the $B_s$-$\overline{B_s}$ mixing.
An interesting point of this setup is that 
the EW-neutral scalar can be interpreted as a good DM candidate 
and such large Yukawa couplings predict a large cross section of
DM annihilation moderately to explain the DM relic density.
Besides, the cross section of the DM direct detection is 
predicted to be just below the current experimental upper bound in this kind of model \cite{Okawa}.
In our model motivated by the excesses in the LHCb experiment,
we have found that the cross section for the DM direct detection has a non-trivial structure:
the photon exchanging and Z exchanging diagrams cancel each other in some parameter region.
Then, the cross section is estimated as the one
just below the current experimental bound \cite{LUX2016,XENON1T} in the region with the $\mathcal{O}(1)$-TeV mass of $\lv$ and about $500$-GeV mass of the DM. 
Around the region, the predicted $C_9$ and $C_{10}$ become large enough to
achieve the $R_K^{(*)}$ experimental results within $1 \sigma$ level.
We can expect that the DM direct detection experiments \cite{XENON1T} will prove our model if our DM is the dominant component of the relic density. 

We can consider the other setups to explain the excesses; for instance, DM is a fermion and extra quarks/leptons are scalar fields. In such a case, large $\lambda_\mu$, which is favored by the excesses, could not be consistent with the
relic density of DM. As studied in Ref. \cite{Okawa},  the relic density requires $|\lambda_\mu| \leq 1$, when
$m_X=500$ GeV and $m_{Q^\prime}=m_{\lv}=1$ TeV. In addition, the constraint from the indirect detection of DM
excludes $m_X \lesssim 1$ TeV in this fermionic DM case. Thus, the scalar DM scenario seems to be favored from
the viewpoint of the consistency between the excesses and the DM observables.

Finally, let us also give a comment on the difference between our setup and the other models that have been studied before. 
In Refs. \cite{Gripaios:2015gra,Arnan:2016cpy}, the explanation of the excesses
has been done using the box diagram involving extra scalars and fermions.
The possible EW charge assignments for the extra fields are well summarized, and
the model with the DM charged under the EW symmetry is discussed in Ref. \cite{Gripaios:2015gra}.
In our paper, we concentrate on the case that the DM is neutral under the EW symmetry, and
we have carefully studied the dark matter physics and collider physics taking into account 
the flavor dependence as well. 
We conclude that both the DM direct detection and
the LHC results strictly limit the explanation of the excess even in our model.
If the DM is charged under the EW symmetry, as discussed in Ref. \cite{Gripaios:2015gra},
the $Z$-boson exchanging would drastically enhance the DM direct-detection cross section
and the bound would become more severe than ours.
We assign the global $U(1)_X$ symmetry to stabilize the DM. 
We can consider the gauge $U(1)_X$ as discussed in Ref. \cite{Belanger:2015nma}.
The authors in Ref. \cite{Belanger:2015nma} consider the possibility that
there is a mass mixing between the extra quark and the SM quarks.
Then, the massive $U(1)_X$ gauge boson exchanging explains the excess. 
In this case, the collider signal of the extra quark
is different from ours, so that the model with the gauged $U(1)_X$ may be able to evade the strong bound from the LHC experiment. Even in such a case, however, we emphasize that
the DM direct detection cross section plays a important role in testing this kind of model,
as far as the DM is complex scalar.
In Ref. \cite{Belanger:2015nma}, the real scalar DM scenario is discussed, so that
the bound from the DM direct detection could be relaxed, although
the Yukawa coupling required by the thermal relic density is larger than the one in our model.


\section*{Acknowledgments}
The work of J. K. is supported by Grant-in-Aid for
Research Fellow of Japan Society for the Promotion of
Science No. 16J04215. The work of Y. O. is supported by Grant-in-Aid for Scientific research from the Ministry of Education, Science, Sports, and Culture (MEXT), Japan, No. 17H05404.

\end{document}